\documentclass[pre,twocolumn,showpacs]{revtex4}
\usepackage{graphicx}
\usepackage{epsfig}
\usepackage{amssymb,amsmath}
\usepackage{bm}

\def\Itens{\mbox{\boldmath$I$\unboldmath}}

\def\qbold{\mbox{\boldmath $q$\unboldmath}}
\def\Dbold{\mbox{\boldmath $D$\unboldmath}}

\def\rbold{\mbox{\boldmath $\hat{r}$\unboldmath}}

\def\ubold{\mbox{\boldmath $u$\unboldmath}}

\def\3dots{\:\raisebox{-0.5ex}{$\stackrel{\textstyle.}{:}$}\:}
\def\beq{\begin{equation}}
\def\eeq{\end{equation}}
\def\bea{\begin{eqnarray}}
\def\eea{\end{eqnarray}}

\begin{document}
\title{Effective medium theory of semiflexible filamentous networks}
\author{Moumita Das$^1$ \footnote{Present address: Department of Physics and Astronomy,
Vrije Universiteit,
Amsterdam, The Netherlands.}}
\author{F. C. MacKintosh $^2$}
\author{Alex J. Levine$^{1,3}$}
\affiliation{$^1$ Department of Chemistry, University of California,
Los Angeles, CA 90095.\\
$^2$ Department of Physics and Astronomy, Vrije Universiteit,
Amsterdam, The Netherlands. \\ $^3$ California Nanosystems
Institute, University of California, Los Angeles, CA 90095. }

\begin{abstract}
We develop an effective medium approach to the
mechanics of disordered, semiflexible polymer networks and study the
response of such networks to uniform and nonuniform strain. We identify
distinct elastic regimes in which the contributions of either filament
bending or stretching to the macroscopic modulus vanish. We also show
that our effective medium theory predicts a crossover between affine and
non-affine strain, consistent with both prior numerical studies and
scaling theory.
\end{abstract}

\date{\today}


\maketitle

Semiflexible polymer networks form a distinct class of gels whose
mechanical properties are important for both biophysical and
materials research. These cross-linked polymer networks differ substantially  from
the flexible polymer gels and rubbers~\cite{rubenstein:03} due to the rigidity of the
individual polymers~\cite{janmey:90,mackintosh:97}.
Because the thermal persistence length of the
constituent filaments is much longer than the typical distance
between cross-links, these materials can store elastic strain energy
in both stretching and bending deformations of the filaments.
The cytoskeleton, a complex assembly that includes stiff filamentous
proteins present in most eukaryotic cells, is an especially common
example of such a semiflexible network~\cite{alberts:88}. Such networks
dominate the mechanical properties of the cytosol and are at the heart of
cellular force production and morphological control.

Theoretical studies of the elastic response of
randomly cross-linked, stiff, filamentous networks have recently uncovered
a surprising cross-over between distinct mechanical regimes of  these
semiflexible networks~\cite{head:03,wilhelm:03}.  For given filament
elastic parameters there is a transition from strain energy storage in
filament stretching modes at higher network densities to filament
bending modes in more sparse networks.
This transition is accompanied by a change in the geometry of the
deformation field over mesoscopic lengths. At higher densities the
network deforms affinely as expected from continuum elasticity
theory while at lower densities, where the elastic
energy is stored in bending modes, the network deformation field is
nonaffine over mesoscopic distances. Recent
experiments~\cite{gardel:06,liu:07} support the existence of this affine (A)
to nonaffine (NA) cross-over. However, a fundamental understanding of the
relation of the network architecture and individual filament
mechanics to the collective elasticity of the network remains
elusive and prior theoretical work has been primarily numerical.

In this letter we develop an analytical model of the mechanical
response of two-dimensional, disordered, semiflexible networks. We
introduce a mechanical mean-field or effective medium theory of the
system that allows us to calculate the elastic response of the
system to uniformly imposed as well as wavenumber-dependent strain
fields. From this mechanical response we identify an $\mbox{A}/\mbox{NA}$
cross-over and obtain a phase diagram of the system showing the
regimes of affine and non-affine behavior. Our study also demonstrates the presence of a
natural length scale controlling the A/NA cross-over that corresponds
well with prior results from simulation and scaling theory~\cite{head:03}.

\begin{figure}
\includegraphics[width=7cm,height=5.5cm]{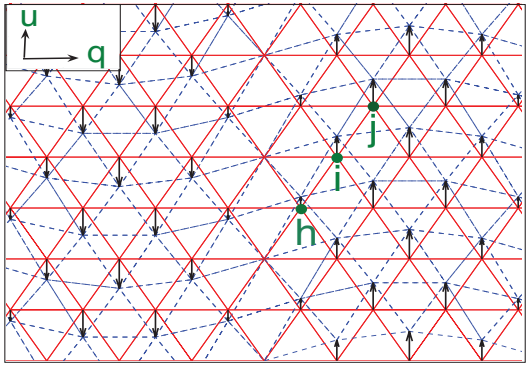}
\caption{\label{schematic}(color online)  Schematic figure of the
filament network. The solid red lines represent the undeformed
filament network, while the dashed blue lines show the
deformation field having wavevector ${\bf q}$ and displacement amplitude
${\bf u}$ (shown in the upper left corner of the figure). The black arrows
show the displacement field at each lattice point.  This perfect
lattice is disordered by making randomly placed cuts in the
infinitely long filaments. These are not shown.}
\end{figure}

We study a model two-dimensional system constructed as follows. We
arrange infinitely long filaments in the plane of a two-dimensional
hexagonal lattice so that at each lattice point three filaments
cross. In this way each lattice point is connected to its nearest
neighbor by a single filament. A sketch of the network is shown
 in Fig.~\ref{schematic}. The filaments are given an
extensional spring constant $\alpha$ and a bending modulus
$\kappa$.  The cross-links at each lattice site do not constrain
the angle between the crossing filaments. We introduce 
finite filament length $L$ into the system by cutting bonds with probability
$1-p$, where $0 < p < 1 $, with no spatial correlations between these cutting points. 
This generates a broad distribution of filament
lengths with a finite  average $\langle L \rangle$, while introducing
quenched disorder. We then study the 
mechanical response of this disordered network in the linear response
regime. The main assumption in our theory is that the depleted network has the same mechanical response as a uniform network with effective elastic constants $\alpha_m$ and $\kappa_m$. These are determined by requiring that strain fluctuations produced in the original, ordered network by randomly cutting the filaments have zero average.
Here, we do not explicitly consider thermal fluctuations, whose
role in determining the longitudinal compliance of filaments has been
discussed before \cite{mackintosh:95}. These thermal effects can be
incorporated in the present model through a renormalized parameter  $\alpha$
\cite{head:03}.

The elastic energy of the strained network, arising from 
bending and stretching of the constituent filaments, can be written
in terms of the displacement vector ${\bf u}_i$ at each lattice site
$i$.  To quadratic order in ${\bf u}$  the stretching ($E_s$) and
bending ($E_b$) energies are
\beq
\label{smalldeformstretching}
E_s  =  \frac{1}{2} \alpha \sum_{\langle ij \rangle}
\left (\ubold_{ij} . \rbold_{ij} \right)^2
\eeq 
\vskip -0.5cm
\beq
\label{smalldeformbending}
E_b =  \frac{1}{2} \kappa R^{-2} \sum_{\langle
\widehat{hij} \rangle} \left( \ubold_{ih} \times \rbold_{ij}
-   \ubold_{ij} \times \rbold_{ih} \right )^2,
\eeq
where $R$ is the lattice constant, ${\rbold}_{ij}$ is a
unit vector directed from the $i^{\mbox{th}}$ to the $j^{\mbox{th}}$
equilibrium lattice site, and ${\bf u}_{ij}$ is the difference in
the strain field between those lattice sites.

It is now simple to determine the collective elastic properties of
the perfect lattice; doing so for the disordered lattice generated
by randomly cutting the filaments is less trivial. We determine
the spring constant and bending modulus of  a spatially
uniform effective system~\cite{Feng} that reproduces the mechanics of
our disordered system in an average sense as described below.

We first apply a uniform dilation to the uniform system with
spring constant $\alpha_m$ so that all bonds
are stretched by ${\delta \ell}$. There is no bending deformation.
If we now replace a single filament segment connecting points (say)
$i$ and $j$ (see Fig.~\ref{schematic}) by one of spring constant
$\alpha^{\prime}$, the virtual force needed to the fix positions of $i$ and $j$
is $f={\delta \ell} (\alpha_m - \alpha^{\prime})$.  If $f$ were applied to
the same segment in the unstrained network the resulting change in length
would be $f/( \alpha_m/a^{*}
-\alpha_m + \alpha^{\prime})$, where $a^{*}$ ($0 <a^{*} <1$) is a network
material parameter that includes the contribution of the elasticity
of the entire network. It may be written in term of the dynamical
matrix of the lattice $\Dbold(q)$ as
\begin{equation}
\label{a-star} a^{*} = \frac{1}{3} \sum_{q} Tr  \left[ \Dbold_s (q)
\cdot \Dbold^{-1} (q) \right ]
\end{equation}
where the sum is over the first Brillouin zone.  Here 
$\Dbold(q)= \Dbold_s (q) + \Dbold_b(q)$, where $\Dbold_{s,b}(q)$ define  the
stretching and bending contributions, respectively, to the full dynamical matrix and
are given by:
\begin{eqnarray}
\label{Ds}
\Dbold_s (q) &=& \alpha_m \sum_{\langle ij \rangle}
\left[ 1 - e^{-i \qbold.\rbold_{ij}} \right]  \rbold_{ij}  \rbold_{ij}  \\
{\Dbold_b}(q) &=& \kappa_m R^{-2} \sum_{\langle ij \rangle}   \left[ 4(1 - \cos(
\qbold.\rbold_{ij}))  \right.  \nonumber  \\  &&  \left. - ( 1 - \cos(
2 \qbold.\rbold_{ij})) \right]   \left(\Itens-
\rbold_{ij}  \rbold_{ij} \right)
\label{Db}
\end{eqnarray}
with $\Itens$ the unit tensor and the sums are over nearest neighbors ~\cite{Feng}.
Note that for small $q$, $\Dbold_b \sim q^4$ and $\Dbold_s \sim q^2$
have the expected wavenumber dependencies for bending and stretching.

From linearity it follows that the extra displacement
${\delta u}$ of the segment $i j$ due to the change in that filament segment's
spring constant in the dilated network is the same as its extension
in response to the force $f$ being applied to it. Therefore 
this additional displacement or `fluctuation' is:
\begin{equation}
\label{delta-u}
\delta u= \frac{(\alpha_m - \alpha^{\prime})
\delta \ell}{ \alpha_m/a^{*} -\alpha_m + \alpha^{\prime}}.
\end{equation}
We now average this extra displacement over the ensemble of possible
filament substitutions, with the statistical distribution of longitudinal spring constants:
\begin{equation}
\label{alpha-dist}
P(\alpha^{\prime})= p \delta(\alpha-\alpha^{\prime}) + (1-p) \delta(\alpha^{\prime}),
\end{equation}
where $1-p$ is the probability of a cut bond and $\delta(\ldots)$ is the Dirac delta function.
To determine the elastic properties of the effective medium we
adjust the medium spring constants $\alpha_m$ so that $\langle
\delta u \rangle = 0$, \emph{i.e.} the lattice displacement in our
spatially homogeneous effective medium material is identical to the
average displacement in the spatially heterogeneous disordered
material.

Using this procedure we arrive at a spatially homogeneous effective medium having
spring constant $\alpha_m$  given by
\begin{equation}
\frac{\alpha_m}{\alpha}= \begin{cases} \frac{p-a^{*}}{1-a^{*}} &
\text{if  $p > a^{*}$,} \\ 0 &  \text{if $p \le a^{*}$.} \end{cases}
\end{equation}
The contribution of network bending to the effective medium spring constant
(which is proportional to the collective shear modulus)
arises only through the effect of the bending modulus on $a^{*}$ in
Eqs.~(\ref{a-star},\ref{Ds},\ref{Db}).
\begin{figure}
\includegraphics[width=8cm,height=5.5cm]{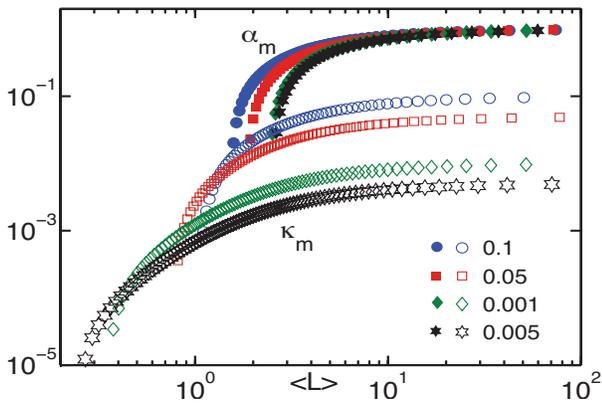}
\caption{\label{effectivemoduliI}(color online) The effective medium
spring constant $\alpha_m$ (filled symbols) and bending constant
$\kappa_m$ (open symbols) with the  average filament length $\langle
L \rangle$   (legend shows different values of $\kappa$, with
$\alpha=1$).}
\end{figure}
To determine how the shear modulus depends on the average filament
length we note the mean filament length is $\langle L \rangle = p R
(2-p)/(1-p)$. We plot in Fig.~\ref{effectivemoduliI} using the
filled symbols the effective medium spring constant as a function of
mean filament length measured in units of $R$.

We now consider the response of the network to a $\qbold$-dependent
strain as depicted in Fig. \ref{schematic}. We modify both the bending modulus
and spring constant of one filament spanning lattice sites $h,i,j$ so
that: $\kappa_m \rightarrow \kappa^{\prime}, \alpha_m \rightarrow \alpha^{\prime}$ and
compute the virtual force and torque needed to maintain the position
of site $i$ in the middle of this triad of lattice sites (Fig.~\ref{schematic}).   
Using these forces and linearity,
we compute the displacement of the $i^{\rm th}$ site in response to the
elastic constant subsitution made above.
We find the displacements along the filament ($\delta \ell_{||}$), and
perpendicular to it ($\delta \ell_{\perp}$) are given by
\begin{eqnarray}
\label{par-displ}
{\delta \ell}_{\parallel}& = &\frac{(\alpha_m - \alpha^{\prime})(\ubold_{ij}+\ubold_{ih}).\rbold_{ij} }{2(\alpha_m/a^{*} - \alpha_m + \alpha^{\prime})},\nonumber \\
\, {\delta \ell}_{\perp}& = & \frac{(\kappa_m - \kappa^{\prime})(\ubold_{ij} + \ubold_{ih}).(\hat{z} \times
\rbold_{ij})}{\kappa_m/b^{*} - \kappa_m + \kappa^{\prime}}
\end{eqnarray}
where $a^*$ is defined in Eq.~(\ref{a-star}) and the analogous quantity $b^*$ is defined by
\begin{equation}
 b^{*} = \frac{1}{3} \sum_{q} Tr  \left[ \Dbold_b (q) \Dbold^{-1} (q) \right]
\end{equation}
using the same sum over wavevectors as in Eq.~(\ref{a-star}). Here, for
semiflexible filaments on a triangular lattice interacting via
cross-links that do not apply torques, the stretching and bending
modes are orthogonal \footnote{For larger deformations, however, the
bending force does have a component along the bond
$\rbold_{ij}$ given by $\frac{\kappa}{R^2}
[(\ubold_{+}.\rbold_{ij})(\ubold_{-}. \rbold_{ij}) - \ubold_{+}.
\ubold_{-}]$ where $\ubold_{+}=\ubold_{ij}+\ubold_{ih}$ and
$\ubold_{-}=\ubold_{ij} - \ubold_{ih}$.}.

We now average these displacements over the disorder, and,
 to find the effective medium elastic constants, we demand that the
disorder-averaged displacements $\langle {\delta \ell}_{\parallel} \rangle$ and
$\langle {\delta \ell}_{\perp} \rangle$ vanish.
The probability distribution for $\alpha^{\prime}$ is given by
Eq.~(\ref{alpha-dist}), but a nonzero value of the bending modulus at
site $i$ requires the presence of \emph{both} filament segments on
either side of that site so that
\begin{equation}
\label{kappa-dist}
P(\kappa^{\prime})= p^2 \delta(\kappa-\kappa^{\prime}) + (1-p^2) \delta(\kappa^{\prime}).
\end{equation}

Since we consider uncorrelated distributions of the bending and
elastic constants we find the effective medium elastic constants
$\alpha_m$ and $\kappa_m$ by solving Eq.~(\ref{par-displ}) for
$\langle {\delta \ell}_{\parallel} \rangle = \langle {\delta \ell}_{\perp} \rangle =0$
independently  to arrive at
\begin{eqnarray}
\frac{\alpha_m}{\alpha} &=& \begin{cases} \frac{p - a^{*}}{1-a^{*}} &
\text{if  $p > a^{*}$} \\ 0 &  \text{if $p \le a^{*}$} \end{cases}\\
\frac{\kappa_m}{\kappa} &=& \begin{cases} \frac{p^2-b^{*}}{1-b^{*}} &
\text{if  $p >  \sqrt{b^{*}}$,} \\ 0 & \text{if $p \le \sqrt{b^{*}}$} \end{cases}.
\end{eqnarray}

Figure \ref{effectivemoduliI} shows the effective medium values of
$\alpha_m$ (filled symbols) and $\kappa_m$ (open symbols) as a
function of  $\langle L \rangle$  for
different values of bending rigidity $\kappa$, at 
fixed value of $\alpha=1$. The
unit of length is the lattice constant $R$ (set to unity) and the
energy scale is arbitrary.

\begin{figure}
\includegraphics[width=8cm]{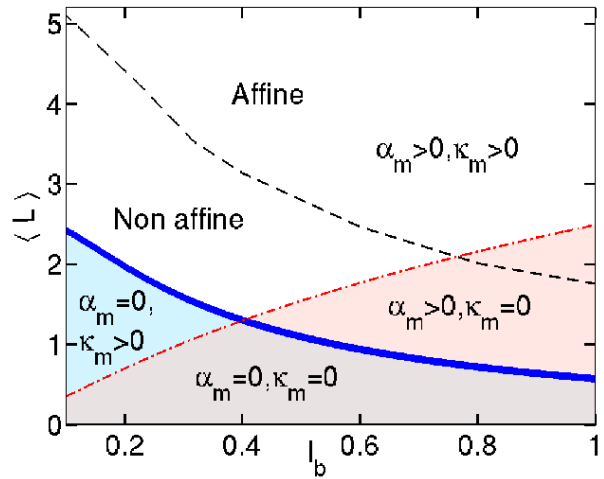}
\caption{\label{phasedig}(color online) The effective medium
mechanical phase diagram spanned by  $\langle L \rangle$ and $l_b$.
The thick solid line marks the rigidity percolation transition where
the material acquires a finite shear modulus. The dashed line
shows the crossover from the non-affine to the affine regime.}
\end{figure}

There are three length scales in the system: (i) the average length
of filaments $\langle L \rangle$, (ii) a length
$l_b=(\kappa/\alpha)^{1/2}$ associated with the relative ease of
filament stretching to bending, and (iii) the mean distance between
cross links, which, to a good approximation is $R$\footnote{The
mean distance between cross links is always slightly greater than
$R$ due to missing cross-linking filaments in sparse networks.
Corrections to the mean distance between cross-links grow as
$(1-p)^4$.}. We present a mechanical phase diagram of our system
spanned by $\langle L \rangle $ and $l_b$ in Fig.\ref{phasedig} that
shows the regimes corresponding to zero and finite values of
$\kappa_m$ and $\alpha_m$. Generically, for long enough filaments the
system has finite collective extension and bending moduli. As the
mean filament length is reduced the collective shear modulus
vanishes at the rigidity percolation
transition~\cite{Feng,latva-kokko:01,head:03a}.  We also find a new rigid phase ($\alpha_m >
0)$ that has a vanishing bending modulus $\kappa_m$. We further note
that the lower range in $l_b$ corresponds to non-thermal systems in
which the distance between crosslinks  is large compared with the
molecular scale (i.e., at low volume fraction),  but that with thermal effects,
one effectively goes to higher $l_b$.

The collective elastic properties of the effective medium can be
calculated from the stored elastic energy density $\mathcal{E}$
under a given imposed network strain. For strain field of the form
$\ubold=  R \gamma \cos(q \cdot {\bf x}) \hat{z} \times \hat{q}$ the
shear $G_{\rm eff}$ and bending moduli $K_{\rm eff}$ can be extracted as the
coefficients of the $q^2$ and $q^4$ terms of $ \langle
\mathcal{E}\rangle /\gamma^2$ where the angled brackets imply an
angular average of the direction of ${\bf q}$ with respect to
underlying lattice. $G_{\rm eff}$ is proportional to
$\alpha_m$ alone while $K_{\rm eff}$ is a function of $\kappa_m$ and
$\alpha_m$.

\begin{figure}
\includegraphics[width=8.5cm,height=5.2cm]{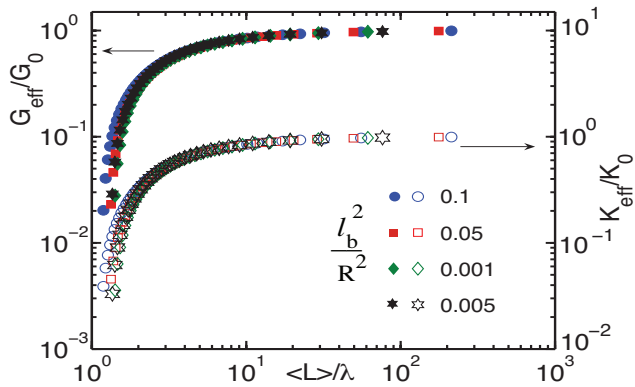}
\caption{\label{effectivemoduliII}(color online)  The effective
medium shear $G_{\rm eff}/G_0$ (filled symbols) and bending moduli
$K_{\rm eff}/K_0$ (open symbols) normalized by their respective values
for a perfect network are plotted as a function of the mean
filament length $\langle L \rangle$  divided by the
nonaffinity length $\lambda=R/{(R/l_b)^z}$ with  $z =0.25$. Data
collapse is shown for four data sets differing in $l_b$. The legend shows
different values of $l_b^2$.}
\end{figure}

In Fig.~\ref{effectivemoduliII} we plot the effective medium shear
and bending moduli as a function of $\langle L \rangle$. Motivated
by earlier work\cite{head:03} on the $\mbox{A}/ \mbox{NA}$ transition we
have rescaled $\langle L \rangle$ by $\lambda = R (R/l_b)^z$ with
$z=1/4$. A comparison of $G_{\rm eff}$  in this figure with
$\alpha_m$($\propto G_{\rm eff}$) from Fig.~\ref{effectivemoduliI} demonstrates a
remarkably accurate data collapse whose accuracy is enhanced as we
move farther away from rigidity percolation. Moreover, we find
that the same rescaling factor generates an equally accurate
collapse of the $K_{\rm eff}$ data.

The collapse of our calculated elastic moduli with a single parameter,
the lengthscale $\lambda$, is in good accord with the numerical data
collapse  found in previous simulations \cite{head:03}. Thus,
the mean-field theory demonstrates all previously observed mechanical
signatures of the $\mbox{A}/ \mbox{NA}$ cross-over seen there.
The analytic results, however,
suggest $z \approx 1/4$, while the prior numerics
pointed to $z=1/3$.  The effective medium approach
introduced here does not allow us to explore the spatial
heterogeneities of the strain field under uniformly imposed shear, so it
is impossible to explore the geometric interpretation of $\lambda$ with this
technique.

Although the effective medium approach fails
to account for the correct spatial structure of the strain field in the
disordered material, it does show an abrupt cross-over that appears
mechanically identical to the $\mbox{A}/ \mbox{NA}$
cross-over. The cross-over is controlled by a single emergent length scale,
$\lambda$, which obeys a similar scaling relation to that found
empirically from previous numerical results. From these $G_{\rm eff}$
plots (Fig.~\ref{effectivemoduliII}) we have extracted the $\mbox{A}/\mbox{NA}$
cross-over from the location of the largest change in the slope of the curves.
This $\mbox{A}/\mbox{NA}$ boundary is plotted in Fig.~\ref{phasedig}.

In conclusion, we used an effective medium theory to explore the mechanical
properties of disordered filament networks. We find that  this
mean-field approach to the mechanics of such networks captures the
mechanical aspects of the A/NA cross-over including the
identification of an emergent mesoscopic length scale $\lambda$
controlling the mechanics of the system.

MD and AJL thank M. F. Thorpe and T.C. Lubensky for useful discussions and acknowledge
support from NSF-DMR0354113. FCM acknowledges the hospitality of the
UCLA chemistry department. This  work was supported in part by the
Foundation for Fundamental Research on Matter (FOM).

\end{document}